\documentclass[graybox]{svmult}
\usepackage[utf8]{inputenc}

\title{HLRS2023}
\author{xu  chu }

\usepackage{float}
\usepackage{mathptmx}       
\usepackage{helvet}         
\usepackage{courier}        
\usepackage{type1cm}        

\usepackage{makeidx}         
\usepackage{graphicx}        
\usepackage{subfigure}
\usepackage{caption}
\usepackage{subcaption}
\usepackage{multicol}        
\usepackage[bottom]{footmisc}
\usepackage{multirow}
\usepackage{latexsym}
\usepackage{epstopdf}
\usepackage{authblk}
\usepackage{amsmath,bm}
\usepackage{amsmath}
\usepackage{amssymb}
\usepackage{tikz}
\usepackage{pgfplots}
\pgfplotsset{compat=1.17}

\makeindex             

\begin{document}

\title*{JAX-based differentiable fluid dynamics on GPU and end-to-end optimization}
\titlerunning{Differentiable fluid dynamics}


\author{Wenkang Wang \inst{1} \and  Xuanwei Zhang \inst{2}   \and Deniz Bezgin \inst{3} \and Aaron Buhendwa \inst{3} \and  Xu Chu \inst{2,4}  \and Bernhard Weigand \inst{5}}

\institute{\inst{1} Max Planck Institute for Intelligent Systems, Heisenbergstraße 3, 70569, Stuttgart, Germany \and %
\inst{2} Cluster of Excellence SimTech, University of Stuttgart, Pfaffenwaldring 5a, 70569 Stuttgart, Germany,  \email{xu.chu@simtech.uni-stuttgart.de} \and %
                    \inst{3} Department of Engineering, University of Exeter, UK \and
                    \inst{4} Chair of Aerodynamics and Fluid Mechanics \and                    
                    \inst{5} Institute of Aerospace Thermodynamics (ITLR), Pfaffenwaldring 31, 70569 Stuttgart, Germany  
                      }
\authorrunning{Wang, Zhang, Bezgin, Buhendwa, Chu, Weigand}

\maketitle
\vspace{-2cm}
\abstract {
This project aims to advance differentiable fluid dynamics for hypersonic coupled flow over porous media, demonstrating the potential of automatic differentiation (AD)-based optimization for end-to-end solutions. Leveraging AD efficiently handles high-dimensional optimization problems, offering a flexible alternative to traditional methods. We utilized JAX-Fluids, a newly developed solver based on the JAX framework, which combines autograd and TensorFlow's XLA. Compiled on a HAWK-AI node with NVIDIA A100 GPU, JAX-Fluids showed computational performance comparable to other high-order codes like FLEXI. Validation with a compressible turbulent channel flow DNS case showed excellent agreement, and a new boundary condition for modeling porous media was successfully tested on a laminar boundary layer case. Future steps in our research are anticipated.}

\section{Introduction}

In recent years, machine learning (ML) has revitalized the physical sciences by introducing innovative tools for predicting the behavior of physical systems. Breakthroughs in ML, coupled with rapid advancements in GPU technology, have sparked a growing interest in the application of machine learning techniques.
There has been considerable discussion about the application of machine learning to sciences such as fluid mechanics \cite{Duraisamy.2019,Chang.2018,chu2018computationally,chu2024non,beck2019deep}. 
Until now, most ML models have been optimized offline, i.e., outside of physics simulators. While offline training of ML models is relatively straightforward, this approach has several drawbacks. Firstly, these models often experience a data-distribution mismatch between the data seen during training and at test time. Secondly, they generally do not benefit from a priori knowledge about the dynamics of the underlying partial differential equations (PDEs). Additionally, fluid mechanics solvers are typically complex, written in low-level programming languages like Fortran or C++, and heavily optimized for CPU computations. This contrasts with practices in ML research, where models are typically trained in Python and optimized for GPU usage. Integrating these ML models into existing CFD software frameworks can therefore be a tedious and challenging task.

To address this problem, researchers have developed differentiable CFD frameworks written entirely in Python, enabling end-to-end optimization of ML models. This end-to-end (online) training approach leverages automatic differentiation (AD) throughout entire simulation trajectories. AD is particularly essential to ML-CFD research as it provides gradient information and facilitates the optimization of both existing and novel CFD models.
JAX is a Python library designed for high-performance numerical computations, utilizing XLA to compile and run code on accelerators such as CPUs, GPUs, and TPUs. Originally created as a machine learning library, JAX supports automatic differentiation, making it an ideal tool for integrating ML with CFD.

Focusing solely on learning or emulating properties of flows overlooks a crucial point. The technological revolution underlying the remarkable advances in machine learning offers the potential for solving classical fluid mechanics problems in a different way, without necessarily using machine learning itself, but by leveraging computational infrastructure. 
At the core of this infrastructure is automatic differentiation, a highly efficient technique for computing derivatives of solution trajectories of partial differential equations. This method is particularly advantageous for solving optimization problems with large numbers of parameters while optimizing complex cost functions. 
Its flexibility allows for rapid experimentation with different approaches and formulations, facilitating the discovery of more intuitive or easier-to-optimize versions. This contrasts with the significantly less flexible classical use of adjoint methods. Inverse design of complex flows is notoriously challenging due to the high cost of high-dimensional optimization. Typically, optimization problems are either limited to a few control parameters or use adjoint-based approaches to transform the optimization problem into a boundary value problem. 
The first aim is to demonstrate the potential of \textit{differentiable fluid dynamics}  for end-to-end optimization of hypersonic coupled flow. This work will pave the way for utilizing automatic differentiation (AD) in a broad range of flow optimization applications in the future.

Besides, it is able to achieve a comprehensive understanding, effective control, and high-dimensional multi-objective optimization of hypersonic flow over porous media. Until now, existing knowledge of flow over porous media has been mostly limited to the incompressible regime. Studying compressible high-speed flow over porous media is crucial for applications such as Thermal Protection Systems (TPS). This objective involves high-dimensional multi-objective optimization to optimize drag and thermal effects, using a comprehensive set of input parameters for porous media. By uncovering the underlying physical mechanisms of these interactions, we will develop strategies for controlling flow behavior, reducing drag, and improving thermal management. Ultimately, this research will optimize the performance and efficiency of hypersonic vehicles involving porous media e.g. as a thermal protection system, paving the way for innovative applications in aerospace technology.

Fluid dynamics within porous media \cite{Chu.2018,Chu.2019,Chu.2020} is complemented by flow across their surfaces, where the interaction can passively or actively influence the flow characteristics.  
Many of the industrial examples can be modeled and simplified as a (turbulent) flow over porous media \cite{Bottaro.2019,Terzis.2019}. 
The phenomenon of free flow over a porous medium is prevalent across a wide array of industrial and environmental applications, encompassing catalytic reactors, heat exchangers, fuel cells, and porous river beds. 
In the quest for mastering space, the study of TPS through the lens of compressible turbulent flow over porous media emerges as a tale of scientific ingenuity and strategic necessity \cite{mansour2024flow}. TPS materials is usually characterized by large porosities ($\ge 80\%$). This is designed to achieve superior insulation performance by reducing the through-thickness heat transfer and by producing endothermic chemical processes within the material that blow cool gases through the structure into the boundary layer. And the porous layer has a significant influence to the thermofluidic behavior of the compressible boundary layer flow.\\


The dynamic interactions at the interface between free flow and porous media involve the mutual exchange of mass, momentum, and energy, which collectively shape the system's overall behavior. Understanding these complex interactions requires thorough investigation and comprehension to accurately characterize and model the involved processes. Researchers have made significant progress in studying free flow over porous media through a combination of experimental, theoretical, and numerical approaches. Direct Numerical Simulation (DNS) is particularly advantageous for observing and analyzing turbulent physics within confined spaces \cite{wang2021anassess,wang2022spatial,wang2021information,Chu.2018,Chu.2016, Pandey.2017b}.
By simulating the scale of the turbulent flow down to Kolmogorov length scale, DNS opens up a new frontier in the modeling and design of more effective, reliable thermal protection systems.
Until now, there are only limited research about fully compressible flow coupled with porous media. Chen and Scalo \cite{chen2021effects} investigated high-Mach-number channel flow where the channel walls are defined with porous boundary condition. In their studies, Large Eddy Simulation (LES) was used to understand how the porous walls influence the flow, especially in terms of pressure changes and stress distribution. 
Zhou et al.\cite{zhou2024direct} used a coupled DNS-volume-averaged Navier–Stokes equations (VANS) approach to simulate compressible turbulent channel flows over porous boundaries.
Deolmi and Müller \cite{deolmi2018two} introduced simulation of a compressible flow over a porous surface and an extension to a two-step model order reduction strategy. 
For the incompressible flow regime, Breugem et al. \cite{Breugem.2005} are one of the first to couple DNS in the channel with VANS in the porous media. The existing Kelvin$-$Helmholtz instability is claimed to be responsible for an exchange in momentum between the channel and the permeable wall.  \\


\section{Numerical approach}

JAX-Fluids \cite{bezgin2023jax,bezgin2024jax} is a Python-based, fully-differentiable CFD solver optimized for compressible single- and two-phase flows, extending its capabilities into high-performance computing environments. Its numerical schemes include a high-order Godunov-type finite-volume formulation, positivity-preserving limiters for increased robustness, and support for stretched Cartesian meshes. The solver utilizes JAX primitives for efficient parallelization on GPU and TPU systems, demonstrating stable automatic differentiation gradients over extended integration trajectories. This advanced functionality allows JAX-Fluids to perform DNS of complex flow phenomena with high-order accuracy and computational efficiency. Enhanced two-phase flow modeling is achieved through both level-set sharp-interface and five-equation diffuse-interface models, ensuring versatile application across a range of fluid dynamics problems. The powerful level-set implementation allows the simulation of arbitrary solid boundaries.\\

The JAX-Fluids solver is characterized by its integration of advanced numerical methods, facilitating a smooth combination of machine learning (ML) with computational fluid dynamics (CFD), alongside its capacity for automatic differentiation (AD). AD stands out as a pivotal feature for ML-CFD research, offering essential gradient data that underpins the optimization of both existing and new CFD models. This functionality is instrumental in enhancing model accuracy and fostering innovation within the fluid dynamics research community.\\

\subsection{Governing equations}

The full compressible Navier-Stokes equations of conservative variables are

\begin{equation}
\frac{\partial \mathbf{U}}{\partial t} + \frac{\partial \mathbf{F}^c(\mathbf{U})}{\partial x} + \frac{\partial \mathbf{G}^c(\mathbf{U})}{\partial y} + \frac{\partial \mathbf{H}^c(\mathbf{U})}{\partial z} = \frac{\partial \mathbf{F}^d(\mathbf{U})}{\partial x} + \frac{\partial \mathbf{G}^d(\mathbf{U})}{\partial y} + \frac{\partial \mathbf{H}^d(\mathbf{U})}{\partial z} + \sum_i \mathbf{S}_i(\mathbf{U})
\label{NSE}
\end{equation}

$\mathbf{F}^c$, $\mathbf{G}^c$, and $\mathbf{H}^c$ denote the convective fluxes in $x$-, $y$- and $z$-direction. Analogously, $\mathbf{F}^d$, $\mathbf{G}^d$, and $\mathbf{H}^d$ denote the dissipative fluxes in the three spatial dimensions. The right-hand side is complemented by the sum of all source terms $\sum_i \mathbf{S}_i(\mathbf{U})$.\\

For single-phase flows, the primitive variables are the fluid density $\rho$, the velocity components $u$, $v$, and $w$ (in $x$-,$y$-, and $z$-direction, respectively), and the pressure $p$. $\mathbf{u}=[u,v,w]^T$ is the velocity vector. $E=\rho e+\frac{1}{2}\rho \mathbf{u} \cdot \mathbf{u}$ denotes the total energy of the fluid. The vector of the Conservative variables is given as

\begin{equation}
\mathbf{U} = \begin{bmatrix}
\rho \\
\rho u \\
\rho v \\
\rho w \\
E \\
\end{bmatrix},
\end{equation}

and the convective fluxes are

\begin{equation}
\mathbf{F}^c(\mathbf{U}) = \begin{bmatrix}
\rho u \\
\rho u^2 + p \\
\rho u v \\
\rho u w \\
u(E + p) \\
\end{bmatrix}, \quad
\mathbf{G}^c(\mathbf{U}) = \begin{bmatrix}
\rho v \\
\rho u v \\
\rho v^2 + p \\
\rho v w \\
v(E + p) \\
\end{bmatrix}, \quad
\mathbf{H}^c(\mathbf{U}) = \begin{bmatrix}
\rho w \\
\rho u w \\
\rho v w \\
\rho w^2 + p \\
w(E + p) \\
\end{bmatrix}.
\end{equation}

The dissipative fluxes describe viscous effects and heat conduction.

\begin{equation}
    \mathbf{F}^d(\mathbf{U}) = \begin{bmatrix}
0 \\
\tau_{xx} \\
\tau_{xy} \\
\tau_{xz} \\
\sum u_i \tau_{ix} - q_x \\
\end{bmatrix}, \quad
\mathbf{G}^d(\mathbf{U}) = \begin{bmatrix}
0 \\
\tau_{yx} \\
\tau_{yy} \\
\tau_{yz} \\
\sum u_i \tau_{iy} - q_y \\
\end{bmatrix}, \quad
\mathbf{H}^d(\mathbf{U}) = \begin{bmatrix}
0 \\
\tau_{zx} \\
\tau_{zy} \\
\tau_{zz} \\
\sum u_i \tau_{iz} - q_z \\
\end{bmatrix}
\end{equation}

The viscous stress is given by

\begin{equation}
\tau_{ij} = \mu \left( \frac{\partial u_i}{\partial x_j} + \frac{\partial u_j}{\partial x_i} - \frac{2}{3}\delta_{ij}\frac{\partial u_k}{\partial x_k} \right),
\end{equation}

where $\mu$ is the dynamic viscosity. The energy flux vector $q = [q_1, q_2, q_3]^T$ is expressed via Fourier’s
heat conduction law, $\mathbf{q} = -\lambda \nabla T $, where $\lambda$ is the heat conductivity.\\

The semi-discrete form of the governing equations \ref{NSE} for the cell-averaged state $\bar{\mathbf{U}}_{i,j,k}$ in cell $(i, j, k)$ is given by

\begin{equation}
\begin{aligned}
\frac{d \bar{\mathbf{U}}_{i,j,k}}{dt} &= - \frac{1}{\Delta x_{i,j,k}}
\left[ \left( \mathbf{F}^{c}_{i+\frac{1}{2},j,k} - \mathbf{F}^{d}_{i+\frac{1}{2},j,k} \right) - \left( \mathbf{F}^{c}_{i-\frac{1}{2},j,k} - \mathbf{F}^{d}_{i-\frac{1}{2},j,k} \right) \right] \\
&\quad - \frac{1}{\Delta y_{i,j,k}}
\left[ \left( \mathbf{G}^{c}_{i,j+\frac{1}{2},k} - \mathbf{G}^{d}_{i,j+\frac{1}{2},k} \right) - \left(\mathbf{G}^{c}_{i,j-\frac{1}{2},k} - \mathbf{G}^{d}_{i,j-\frac{1}{2},k} \right) \right] \\
&\quad - \frac{1}{\Delta z_{i,j,k}}
\left[ \left(\mathbf{H}^{c}_{i,j,k+\frac{1}{2}} - \mathbf{H}^{d}_{i,j,k+\frac{1}{2}} \right) - \left( \mathbf{H}^{c}_{i,j,k-\frac{1}{2}} - \mathbf{H}^{d}_{i,j,k-\frac{1}{2}} \right) \right] \\
&\quad + \bar{\mathbf{S}}_{i,j,k} = \mathcal{R}_{\text{NSE}}
\end{aligned}
\label{ODE}
\end{equation}

For the calculation of the convective intercell numerical fluxes, WENO-type high-order discretization schemes in combination with approximate Riemann solvers is used. Dissipative fluxes are calculated by central finite-differences. The utilization of finite-volume discretization results in the formulation of a series of ordinary differential equations (ODEs) (Eq.\ref{ODE}). These equations are subsequently integrated over time using a preferred ODE solver. Commonly, explicit Runge-Kutta methods that are total-variation diminishing (TVD) are employed to ensure stability and accuracy of the solution.
Furthermore, the level-set implementation allows for arbitrary immersed solid boundaries.


\subsection{Parallel strategy}

JAX-Fluids utilize JAX's inherent tools to implement our parallelization strategy, ensuring both differentiability and optimal performance. Specifically, it employs \texttt{jax.pmap} to express single-program multiple-data (SPMD) code, transforming compute-intensive functions. Like \texttt{jax.jit}, \texttt{jax.pmap} compiles the function using the XLA (Accelerated Linear Algebra) compiler and then executes it in parallel across the specified devices. Additionally, JAX-Fluids uses \texttt{jax.lax.ppermute} to perform collective data permutations between devices. This operation requires each device to send data to and receive data from exactly one other device.

For single-device simulations, the computational domain in JAX-Fluids is shaped as \([N_x + 2N_h, N_y + 2N_h, N_z + 2N_h]\), where \(N_i\), \(i \in \{x, y, z\}\), represents the number of cells in the spatial directions, and \(N_h\) is the number of halo cells. JAX-Fluids decomposes the computational domain into an \(S_x \times S_y \times S_z\) grid of equally sized blocks, where \(S_i\), \(i \in \{x, y, z\}\), represents the number of blocks in the spatial directions. A single block has the shape \([N_x/S_x + 2N_h, N_y/S_y + 2N_h, N_z/S_z + 2N_h]\).

For multi-device simulations, the shape of the entire computational domain becomes \([S_T, N_x/S_x + 2N_h, N_y/S_y + 2N_h, N_z/S_z + 2N_h]\), where the leading array axis has length \(S_T = S_x S_y S_z\). 
The transformation \texttt{jax.pmap} maps a function over the leading array axis \(S_T\), replicating the function on each XLA device and executing it in parallel. An XLA device generally refers to any computational device targeted by the XLA compiler. The term XLA device means synonymously with GPUs and TPUs.

\section{Results and Discussion}

\subsection{Compressible turbulent channel flow}

Turbulent channel flow is considered as a classic setups to investigate turbulence that is constrained by walls. The Mach number, \( M_{b} \), is defined as the ratio of the bulk velocity, \( U_b \), to the speed of sound in the wall, \( c_w \), which is given by \( M_{b} = \frac{U_b}{c_w} = 1.5 \). The Reynolds number, \( Re_b \), based on the bulk density, \( \rho_b \), bulk velocity, \( U_b \), the half-width of the channel, \( h \), and the wall viscosity, \( \mu_w \), is expressed as \( Re_b = \frac{\rho_b U_b h}{\mu_w} = 3000 \). The computational grid consists of a discretization using 
$256\times128 \times128$ cells.
The TENO6-A scheme for cell face reconstruction along with the HLLC Riemann solver for flux calculations is used. The TENO6-A reconstruction method is enhanced by an interpolation limiter, which is tailored to refine the solution. For the cases examined in this work, which do not present severe shock discontinuities, the application of flux limiters is deemed unnecessary. When it comes to diffusive fluxes, we employ central finite-difference schemes of sixth-order accuracy to achieve a precise spatial discretization. 

\begin{figure}[htbp]
    \centering
    \includegraphics[height=4.2cm, keepaspectratio]{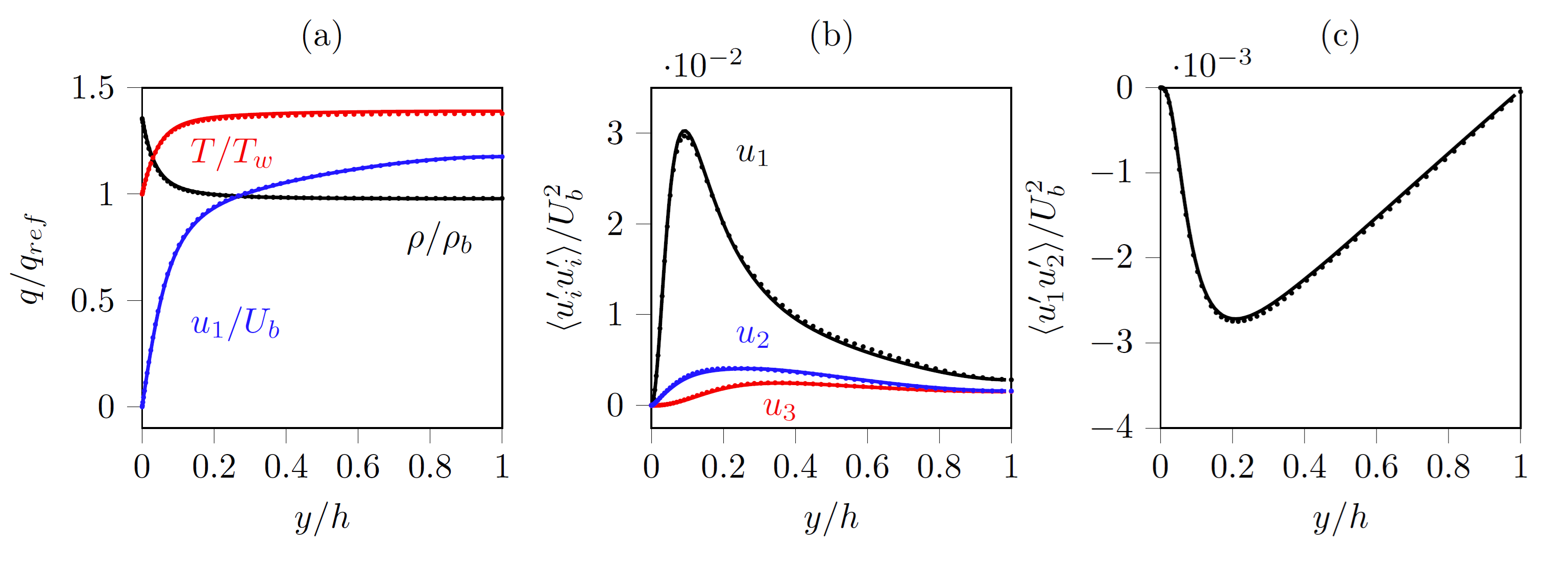}
    \caption{
        Validation of Supersonic Turbulent Channel Flow at $\text{Ma}_b = 1.5$ and $\text{Re}_b = 3000$.
        (a) Normalized mean profiles of density $\frac{\rho}{\rho_b}$, temperature $\frac{T}{T_w}$, and streamwise velocity $\frac{u_1}{U_b}$.    
        (b) Normalized Reynolds normal stresses $\frac{\langle u'_i u'_i \rangle}{U_b^2}$.         
        (c) Normalized Reynolds shear stress $\frac{\langle u'_1 u'_2 \rangle}{U_b^2}$.        
        Reference data is taken from Coleman et al. \cite{coleman1995numerical}. The JAX-Fluids solution is illustrated with solid lines, while the reference by Coleman et al. \cite{coleman1995numerical} is depicted with solid markers.
    }
    \label{fig:validation}
\end{figure}

We conduct a precursor simulation of the channel flow on a coarser computational grid until a statistically steady state is achieved. The simulation starts with a perturbation method (borrowed from perturbU used in OpenFOAM, along with constant pressure and density. Once a statistically steady state is reached, we interpolate the flow field onto the detailed DNS grid. The DNS is executed on eight Nvidia A100 GPUs with single precision. We allow 10 characteristic problem times $h/u_\tau \approx 18.38$ for initial transients to dissipate. Subsequently, we collect 200 snapshots of the instantaneous flow field over a period of approximately $41h/u_\tau$ to compute flow statistics.
Figure \ref{fig:validation} compares the flow statistics to the DNS data from Coleman et al. \cite{coleman1995numerical}. We observe excellent quantitative agreement for mean flow profiles and Reynolds stresses with the cited reference. Similarly, root-mean-square fluctuations of density and temperature (not shown) are in good agreement with the reference data.

\subsection{Compressible laminar boundary layer with wall disturbances }
The JAX-Fluids program implements several widely used boundary conditions, such as no-slip wall, Dirichlet, zero-gradient, etc. The boundary conditions are imposed by assigning values to the halo cells outside of the computation domain. We noticed that in the code of JAX-Fluids, this value assigning process includes the $eval()$ function, which transforms the velocity functions on the boundary into a value. This transformation breaks the chain that links the variable in boundary condition functions to the computational domain. In other words, it is impossible to conduct auto-differentiation with respect to the variables in the boundary condition, which limits the capability of JAX-Fluids on end to end optimization. 

To enable auto-differentiation on boundary condtion variables, we build a new class of boundary condition "Disturbance Wall". Unlike other boundary conditions in JAX-Fluids, which is defined in the input files with anonymous functions, the new boundary condition has its velocity function directly coded in the class, and takes input from an dictionary passed from the top level solver class. Since the form of disturbance velocity on the wall is programmed within the class itself so that the $eval()$ function can be bypassed, which makes it differentiable. We tested the boundary condition on a compressible laminar boundary layer.

\begin{figure}[htbp]
    \centering
    \includegraphics[height=4cm, keepaspectratio]{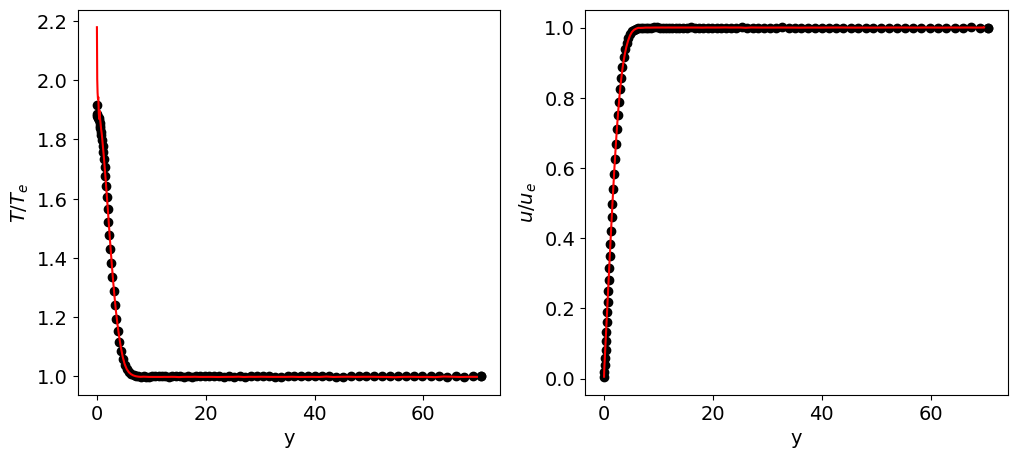}
    \caption{Compressible laminar boundary layer. Normalized temperature $T /T_e$ and normalized  velocity $u/u_e$ profiles
over the self-similar variable $\eta$. The solid lines and markers indicate the JAX-Fluids and Blasius solution, respectively.
}
    \label{fig:laminar1}
\end{figure}

First we set the wall disturbance velocity to zero, which makes the boundary condition exactly the same with a no-slip wall. We use an ideal gas with $\gamma$ = 1.4. The computational domain (x, y) $\in$ [1.0, 1.5] $\times$ [0.0, 0.4] is discretized using 300×200 cells with a uniform grid in x-direction and a stretched mesh in y direction. The stretching parameter is $\beta$ = 2.2. The boundary conditions are a no-slip adiabatic wall at the south boundary and zero-gradient extrapolation at north and east boundaries. At the inlet (west), we impose the self-similar solution, which is computed by solving the compressible Blasius
similarity equations. For the given Mach number $M$ = 2.25 and Prandtl number $Pr$ = 0.72, we get the
normalized velocity $u$/$u_e$ and the temperature $T /T_e$ over the self-similar variable.
\begin{equation}
    \eta=\frac{u_e}{\sqrt{2 \rho_e \mu_e u_e x}} \int_0^y \rho d y.
\end{equation}
The free stream unit Reynolds number is $Re = u_e\rho_e/\mu_e = 10000$. The
temperature dependent dynamic viscosity is computed using the Sutherland law. We evaluate the solution at the outlet boundary
x = 1.5 and present a comparison to the self-similar Blasius solution. Figure \ref{fig:laminar1} depicts the normalized
velocity $u/u_e$ and the temperature $T /T_e$ over the self-similar variable $\eta$. We achieve good agreement with the Blasius solution.

\begin{figure}[htbp]
    \centering
    \includegraphics[height=8cm, keepaspectratio]{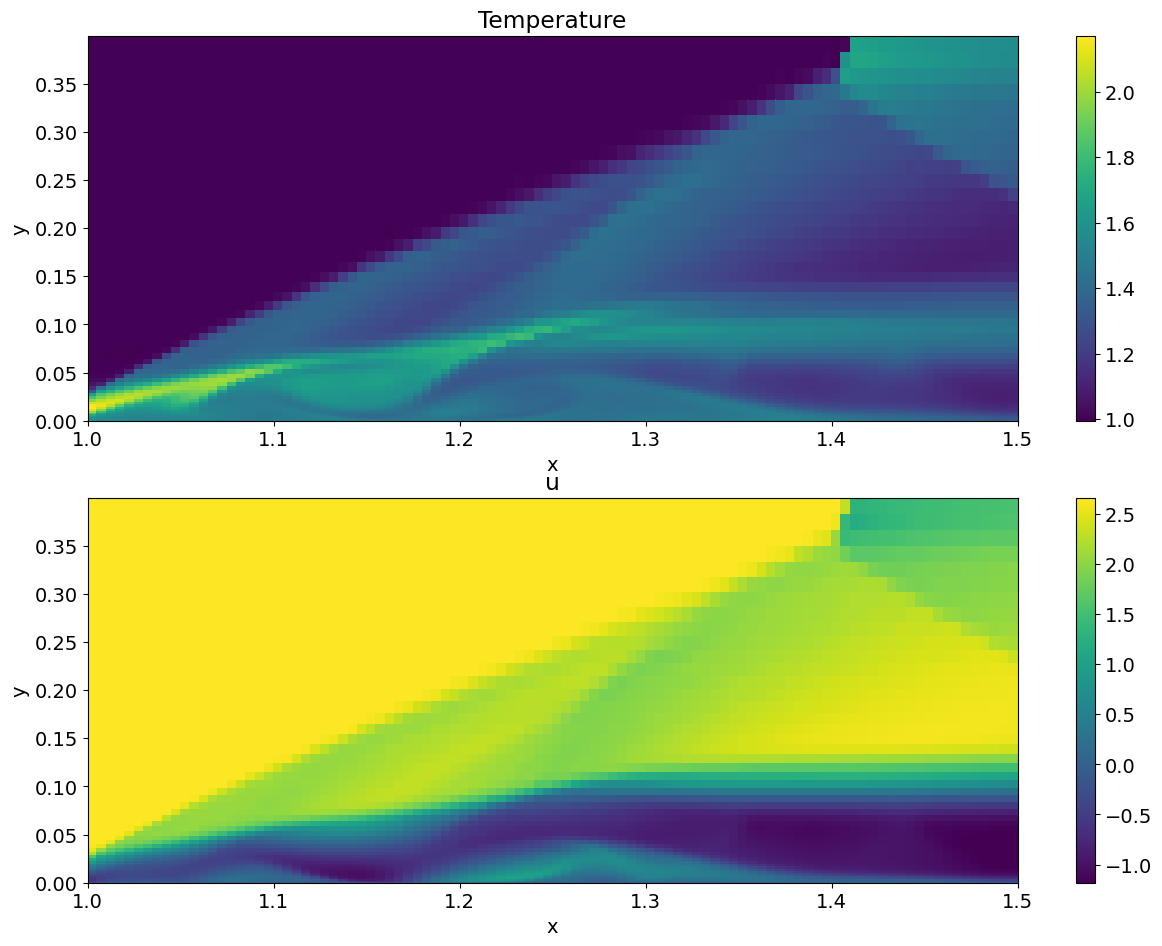}
    \caption{Compressible laminar boundary layer with disturbance wall velocity. The contours show the spatial distribution of temperature and streamwise velocity.
}
    \label{fig:laminar2}
\end{figure}

Then we change the wall normal velocity to a wave-like functions with our customized boundary condition:
\begin{equation}
    v=A\mathrm{sin}(2\pi\kappa x)
\end{equation}
The amplitude A is set to 0.01 and the wavenumber $\kappa=10$. It is noticed that a complex system of shock waves is generated due to the small disturbance on the wall (figure \ref{fig:laminar2}). In future works, we will look deeper into the impact of wall disturbance on the boundary layer and use the end-to-end optimization function to search for disturbance forms that benefit drag reduction and heat prevention.

\section{Parallel performance}
\label{sec:5}

The supercomputing systems used for the DNS is {\it HAWK-AI} located at the High Performance Computer Center Stuttgart (HLRS). The new flagship machine {\it HAWK}, based on the Hewlett Packard Enterprise platform running AMD EPYC 7742 processor code named {\it Rome}, has a theoretical peak performance of 26 petaFLOPs, and consists of a 5,632-node cluster. Such type of GPU-based HPC has been used for previous HPC projects with OpenFOAM \cite{Evrim.2020, Chu.2016, Chu.2016c, Pandey.2017, Pandey.2018, Pandey.2018b, Yang.2020} or with the high-order code Nektar++ \cite{chu2022investigation,Chu.2021b}.
Compared to that, one HAWK-AI node has 8 NVIDIA A100 GPU and two AMD EPYC Rome 7702 CPU processors.
Since now, our simulations will be mostly conducted on the GPU-based nodes which is necessary for the transition to the next-generation HPC Hunter and Herder.

The parallel performance was tested using the DNS of compressible turbulent channel flow case in subsection 3.1. It has a total of 4.2 Mio. cells. Both the averaged clock time per time step (weak scaling) and averaged wall clock time step cell was tested. The averaged wall clock time step cell is also called as performance index (PID) in other publications \cite{kempf2024gal}. It is calculated with 

\begin{equation}
    \mathrm{PID=\frac{Walltime \times \#Ranks}{\#DOF}}
\end{equation}

\begin{figure}[htbp]
\centering
\begin{subfigure}
    \centering
    \includegraphics[height=4cm, keepaspectratio]{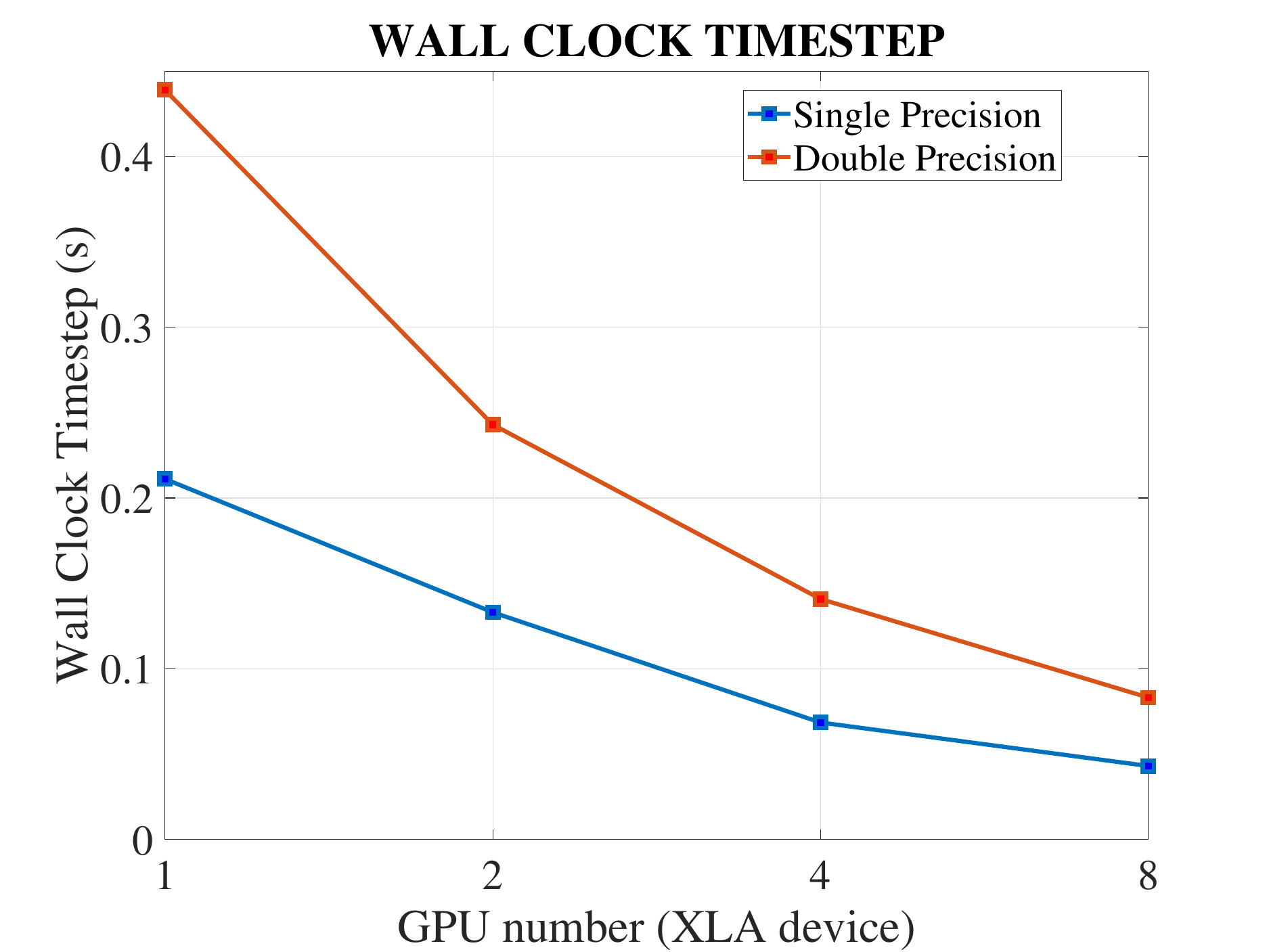}
\end{subfigure}
\begin{subfigure}
     \centering
    \includegraphics[height=4cm, keepaspectratio]{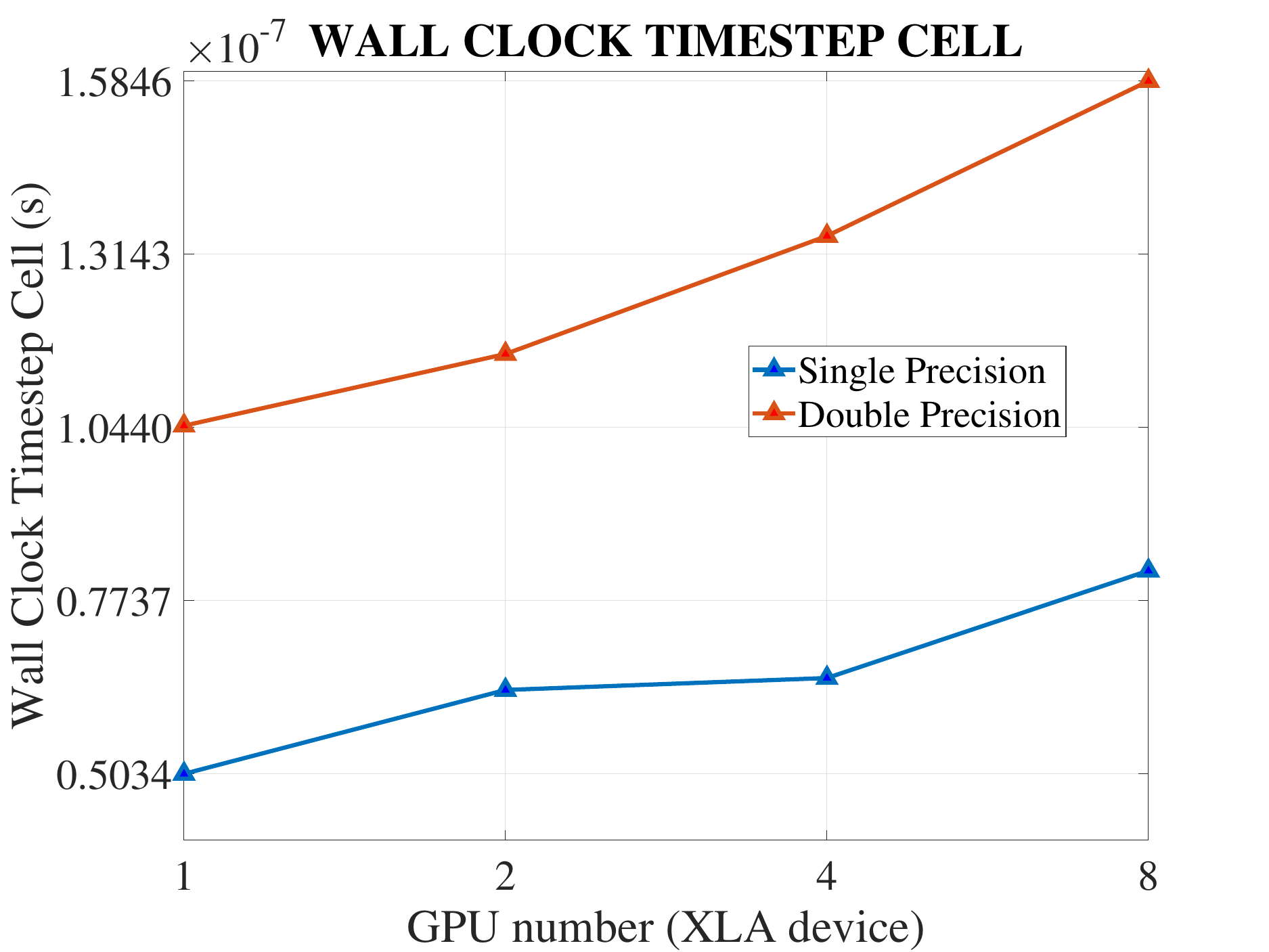}
\end{subfigure}
    \caption{(a): weak scaling, averaged clock time per time step and (b): averaged wall clock time step cell}
    \label{scalability}
\end{figure}

In Figure \ref{scalability}, the wall clock time is lower than 0.1 second per time step with 8 GPUs, corresponding to approximately half Mio. of DOF per GPU. The wall clock time with double precision is about twice of that of single precision.
The The PID of JAX-Fluids (single precision) is $O(10^{-8})$, which is similar to other state-of-the-art discontinuous Galerkin solver GAL\AE XI written in FORTRAN \cite{kempf2024gal}.

Our current scalability test is limited to a single HAWK-AI node due to differences between the PBS scheduler at HLRS and the SLURM scheduler at Jülich. Slight tuning is required to enable multi-node simulations of JAX-Fluids. However, the solver has been tested on a similar system, the JUWELS Booster node, equipped with 4 NVIDIA A100 GPUs, each with 40GB of RAM. Excellent scaling up to 512 A100 GPUs was observed for all models \cite{bezgin2024jax}. The largest simulation, using single-phase simulations with the 512 GPUs, encompasses 16 billion cells.

\section{Conclusions}

This project aims to advance the \textit{differentiable fluid dynamics} on hypersonic coupled flow over porous media. The objective is to demonstrate the potential of automatic differentiation-based optimization for end-to-end optimization. This approach leverages AD to efficiently handle high-dimensional optimization problems, providing a flexible alternative to traditional methods. AD-based optimization has great potential for addressing notoriously challenging problems due to the high cost of high-dimensional optimization.
For this purpose, we utilized a newly developed solver, JAX-Fluids, based on the JAX framework. Google JAX is a machine learning framework for transforming numerical functions in Python. JAX combines a modified version of autograd (automatic differentiation of functions to obtain gradient functions) with TensorFlow's XLA (Accelerated Linear Algebra). It is designed to closely follow the structure and workflow of NumPy and is compatible with various existing frameworks such as TensorFlow and PyTorch.
We compiled the JAX-Fluids solver on a HAWK-AI node equipped with an NVIDIA A100 GPU. Its computational performance (PID) is comparable to other high-order codes like FLEXI on the A100 nodes. The solver has been validated using a compressible turbulent channel flow DNS case, showing excellent agreement. Additionally, we implemented a new boundary condition to model porous media and tested it on a laminar boundary layer case. We are excited about the next steps in our research.

\section{Acknowledgement}
The study has been financially supported by SimTech (EXC 2075/1–390740016) and SFB-1313 (Project No.327154368) from Deutsche Forschungsgemeinschaft (DFG). X. Zhang acknowledges the support from the Chinese Scholarship Council (CSC).

{\footnotesize \bibliography{reference}}

\begin{thebibliography}{10}
\providecommand{\url}[1]{{#1}}
\providecommand{\urlprefix}{URL }
\expandafter\ifx\csname urlstyle\endcsname\relax
  \providecommand{\doi}[1]{DOI~\discretionary{}{}{}#1}\else
  \providecommand{\doi}{DOI~\discretionary{}{}{}\begingroup \urlstyle{rm}\Url}\fi

\bibitem{beck2019deep}
Beck, A., Flad, D., Munz, C.D.: Deep neural networks for data-driven les closure models.
\newblock Journal of Computational Physics \textbf{398}, 108910 (2019)

\bibitem{bezgin2023jax}
Bezgin, D., Buhendwa, A., Adams, N.: Jax-fluids: A fully-differentiable high-order computational fluid dynamics solver for compressible two-phase flows.
\newblock Computer Physics Communications \textbf{282}, 108527 (2023)

\bibitem{bezgin2024jax}
Bezgin, D., Buhendwa, A., Adams, N.: Jax-fluids 2.0: Towards hpc for differentiable cfd of compressible two-phase flows.
\newblock arXiv preprint arXiv:2402.05193  (2024)

\bibitem{Bottaro.2019}
Bottaro, A.: Flow over natural or engineered surfaces: an adjoint homogenization perspective.
\newblock Journal of Fluid Mechanics \textbf{877} (2019)

\bibitem{Breugem.2005}
Breugem, W., Boersma, B.: Direct numerical simulations of turbulent flow over a permeable wall using a direct and a continuum approach.
\newblock Physics of fluids \textbf{17}(2), 025103 (2005)

\bibitem{Chang.2018}
Chang, W., Chu, X., Fareed, A., Pandey, S., Luo, J., Weigand, B., Laurien, E.: Heat transfer prediction of supercritical water with artificial neural networks.
\newblock Applied Thermal Engineering \textbf{131}, 815--824 (2018)

\bibitem{chen2021effects}
Chen, Y., Scalo, C.: Effects of porous walls on near-wall supersonic turbulence.
\newblock Physical Review Fluids \textbf{6}(8), 084607 (2021)

\bibitem{chu2018computationally}
Chu, X., Chang, W., Pandey, S., Luo, J., Weigand, B., Laurien, E.: A computationally light data-driven approach for heat transfer and hydraulic characteristics modeling of supercritical fluids: From dns to dnn.
\newblock International Journal of Heat and Mass Transfer \textbf{123}, 629--636 (2018)

\bibitem{Chu.2016c}
Chu, X., Laurien, E., McEligot, D.M.: Direct numerical simulation of strongly heated air flow in a vertical pipe.
\newblock International Journal of Heat and Mass Transfer \textbf{101}, 1163--1176 (2016)

\bibitem{Chu.2016}
Chu, X., Laurien, E., Pandey, S.: Direct numerical simulation of heated pipe flow with strong property variation.
\newblock In: High Performance Computing in Science and Engineering{\'{}} 16, pp. 473--486. Springer (2016)

\bibitem{Chu.2021b}
Chu, X., M{\"u}ller, J., Weigand, B.: Interface-resolved direct numerical simulation of turbulent flow over porous media.
\newblock In: High Performance Computing in Science and Engineering '19, pp. 343--354. Springer International Publishing (2021)

\bibitem{chu2024non}
Chu, X., Pandey, S.: Non-intrusive, transferable model for coupled turbulent channel-porous media flow based upon neural networks.
\newblock Physics of Fluids \textbf{36}(2), 025112 (2024)

\bibitem{chu2022investigation}
Chu, X., Wang, W., Weigand, B.: An investigation of information flux between turbulent boundary layer and porous medium.
\newblock In: International Conference on High Performance Computing in Science and Engineering, pp. 183--196. Springer (2022)

\bibitem{Chu.2018}
Chu, X., Weigand, B., Vaikuntanathan, V.: Flow turbulence topology in regular porous media: From macroscopic to microscopic scale with direct numerical simulation.
\newblock Physics of Fluids \textbf{30}(6), 065102 (2018)

\bibitem{Chu.2020}
Chu, X., Wu, Y., Rist, U., Weigand, B.: Instability and transition in an elementary porous medium.
\newblock Physical Review Fluids \textbf{5}(4), 044304 (2020)

\bibitem{Chu.2019}
Chu, X., Yang, G., Pandey, S., Weigand, B.: Direct numerical simulation of convective heat transfer in porous media.
\newblock International Journal of Heat and Mass Transfer \textbf{133}, 11--20 (2019)

\bibitem{coleman1995numerical}
Coleman, G., Kim, J., Moser, R.D.: A numerical study of turbulent supersonic isothermal-wall channel flow.
\newblock Journal of Fluid Mechanics \textbf{305}, 159--183 (1995)

\bibitem{deolmi2018two}
Deolmi, G., M{\"u}ller, S.: A two-step model order reduction method to simulate a compressible flow over an extended rough surface.
\newblock Mathematics and Computers in Simulation \textbf{150}, 49--65 (2018)

\bibitem{Duraisamy.2019}
Duraisamy, K., Iaccarino, G., Xiao, H.: Turbulence modeling in the age of data.
\newblock Annual Review of Fluid Mechanics \textbf{51}, 357--377 (2019)

\bibitem{Evrim.2020}
Evrim, C., Chu, X., Laurien, E.: Analysis of thermal mixing characteristics in different t-junction configurations.
\newblock International Journal of Heat and Mass Transfer \textbf{158}, 120019 (2020)

\bibitem{kempf2024gal}
Kempf, D., Kurz, M., Blind, M., Kopper, P., Offenh{\"a}user, P., Schwarz, A., Starr, S., Keim, J., Beck, A.: Gal $\{$$\backslash$AE$\}$ xi: Solving complex compressible flows with high-order discontinuous galerkin methods on accelerator-based systems.
\newblock arXiv preprint arXiv:2404.12703  (2024)

\bibitem{mansour2024flow}
Mansour, N., Panerai, F., Lachaud, J., Magin, T.: Flow mechanics in ablative thermal protection systems.
\newblock Annual Review of Fluid Mechanics \textbf{56}, 549--575 (2024)

\bibitem{Pandey.2017}
Pandey, S., Chu, X., Laurien, E.: Investigation of in-tube cooling of carbon dioxide at supercritical pressure by means of direct numerical simulation.
\newblock International Journal of Heat and Mass Transfer \textbf{114}, 944--957 (2017)

\bibitem{Pandey.2018b}
Pandey, S., Chu, X., Laurien, E.: Numerical analysis of heat transfer during cooling of supercritical fluid by means of direct numerical simulation.
\newblock In: High Performance Computing in Science and Engineering'17, pp. 241--254 (2018)

\bibitem{Pandey.2018}
Pandey, S., Chu, X., Laurien, E., Weigand, B.: Buoyancy induced turbulence modulation in pipe flow at supercritical pressure under cooling conditions.
\newblock Physics of Fluids \textbf{30}(6), 065105 (2018)

\bibitem{Pandey.2017b}
Pandey, S., Laurien, E., Chu, X.: A modified convective heat transfer model for heated pipe flow of supercritical carbon dioxide.
\newblock International Journal of Thermal Sciences \textbf{117}, 227--238 (2017)

\bibitem{Terzis.2019}
Terzis, A., Zarikos, I., Weishaupt, K., Yang, G., Chu, X., Helmig, R., Weigand, B.: Microscopic velocity field measurements inside a regular porous medium adjacent to a low reynolds number channel flow.
\newblock Physics of Fluids \textbf{31}(4), 042001 (2019)

\bibitem{wang2021information}
Wang, W., Chu, X., Lozano-Dur{\'a}n, A., Helmig R.and~Weigand, B.: Information transfer between turbulent boundary layers and porous media.
\newblock Journal of Fluid Mechanics \textbf{920} (2021)

\bibitem{wang2022spatial}
Wang, W., Lozano-Dur{\'a}n, A., Helmig, R., Chu, X.: Spatial and spectral characteristics of information flux between turbulent boundary layers and porous media.
\newblock Journal of Fluid Mechanics \textbf{949}, A16 (2022)

\bibitem{wang2021anassess}
Wang, W., Yang, G., Evrim, C., Terzis, A., Helmig, R., Chu, X.: An assessment of turbulence transportation near regular and random permeable interfaces.
\newblock Physics of Fluids \textbf{33}(11), 115103 (2021).
\newblock \doi{10.1063/5.0069311}

\bibitem{Yang.2020}
Yang, G., Chu, X., Vaikuntanathan, V., Wang, S., Wu, J., Weigand, B., Terzis, A.: Droplet mobilization at the walls of a microfluidic channel.
\newblock Physics of Fluids \textbf{32}(1), 012004 (2020)

\bibitem{zhou2024direct}
Zhou, Z., Huang, W.X., Xu, C.X.: Direct numerical simulation of compressible turbulent channel flows over porous boundaries.
\newblock Physics of Fluids \textbf{36}(5) (2024)

\end{thebibliography}
\bibliographystyle{spmpsci}

\end{document}